\documentclass[twocolumn, usenatbib]{openjournal}
\usepackage{graphicx} % Required for inserting images
\usepackage{multirow}

\usepackage[colorlinks,linkcolor=blue,citecolor=blue,urlcolor=blue ]{hyperref}
\usepackage[utf8]{inputenc}
\usepackage{float}
\usepackage{xcolor}
\usepackage{ulem}
\usepackage[T1]{fontenc}

\def\logm{log (M$_{\star}$/M$_{\odot}$)}
\def\msun{M$_{\odot}$}

\def\sigh2{$\Sigma_{\rm H_2}$}

\def\sigstar{$\Sigma_{\star}$}

\begin{document}
\title{Galaxy evolution in the post-merger regime.  IV - The long-term effect of mergers on galactic stellar mass growth and distribution\vspace{-1.5cm}}
\author{Sara L. Ellison$^1$}
\author{Leonardo Ferreira$^{1,2}$}
\thanks{$^*$E-mail: sarae@uvic.ca}
\affiliation{$^1$ Department of Physics \& Astronomy, University of Victoria, Finnerty Road, Victoria, BC V8P 1A1, Canada\\
  $^2$ Instituto de Matemática, Estatística e Física, Universidade Federal do Rio Grande, Brazil
}

\begin{abstract}

Galaxy mergers are known to trigger bursts of central star formation, which should therefore lead to stellar mass growth in their inner regions.  However, observational measurements of this `burst mass fraction' are scant.  Here, we assemble a large sample of $\sim$ 14,000 post-coalescence galaxies that have recently completed their merger-induced star formation, and compare various measurements of central stellar mass with a matched control sample.  Specifically, we quantify (at fixed redshift, star formation rate and total stellar mass) the stellar mass enhancement within a fixed angular aperture ($\Delta M_{\mathrm{\star,fibre}}$) and in the galactic bulge ($\Delta M_{\mathrm{\star,bulge}}$), finding burst mass fractions of 10 -- 20\%.  61 galaxies in our sample are at $z\le0.05$ and have integral field unit data from the Mapping Galaxies at Apache Point (MaNGA) survey, allowing further kpc-scale assessment of excess stellar mass and radial gradients.  When assessed within apertures defined in units of kpc we again find a $\sim$ 15 -- 20\% excess of stellar mass in the central regions of the post-mergers compared with matched controls. However, within apertures defined in units of effective radius this stellar mass enhancement increases to 40\%, suggesting that the relative structure/size of the galaxy is important for regulating the location of the merger induced star formation.  Moreover, we find that these stellar mass enhancements are spatially extended, out to $\sim$7 kpc or around 1 R/R$_{\mathrm{e}}$, although the small sample size of the MaNGA overlap limits our radial sampling.   Our work represents the first direct measurement of merger-induced stellar mass growth that is independent of stellar population modelling, or fitting light profiles, demonstrating significant and extended mass build-up in late stage post-mergers. 
  
\end{abstract}
\maketitle

\section{Introduction} \label{intro_sec}

Galaxy mergers represent a cornerstone in the contemporary paradigm of hierarchical growth.  The coalescence of two galaxies leads to the natural build-up of all of their constituent parts, from the stars and gas, to their dark matter halos and central supermassive black holes.  Merger histories can be witnessed in nearby galaxies in the form of extended low surface brightness features (usually the remnants of relatively major mergers) and stellar streams that result from satellite accretion events \citep[e.g. ][]{Karademir2019, Valenzuela2024}.  However, galactic growth happens in tandem via \textit{in-situ} processes unrelated to galaxy mergers, for example, as gas is transformed into stars through gravitational collapse in the disk and the central supermassive black hole is fed via secular means.  A pertinent question is therefore: what fraction of galaxy growth occurs from star formation enhancement as a direct result of a merger versus secular \textit{in-situ} processes? 

\medskip

In the current paper, we focus on the specific topic of stellar mass growth.  Several works have attempted to quantify the fraction of \textit{ex-situ} stellar mass in galaxies, i.e., the fraction of stellar mass that has been accreted through a merger, rather than formed within the current host.  Quantifying the \textit{ex-situ} stellar mass fraction is relatively straightforward in simulations, whereby the origin of particles can be readily tracked \citep[e.g.][]
{Kannan15,RodriguezGomez2016, Boecker2023, Cannarozzo2023}.  Observationally, quantifying the \textit{ex-situ} mass fraction is more challenging, but several works have attempted the measurement \citep{Boecker2020, Angeloudi2024, Angeloudi2025, Cai2025}.  The general consensus is that whilst \textit{ex-situ} stellar mass is most readily accreted at larger radii, there is considerable spatial variation due to properties such as mass ratio and dynamics \citep[e.g.][]{RodriguezGomez2016, Karademir2019, Angeloudi2025}. 

\medskip

Previous works attempting to quantify the contribution of mergers to stellar mass growth have generally focussed on stars that previously existed in a companion galaxy but that have been accreted during coalescence.  I.e. mergers are generally considered to be the source of \textit{ex-situ} mass.  However, this assessment overlooks the fact that star formation occurs as a direct result of the galaxy-galaxy interaction.  It is very well documented that galaxy-galaxy interactions lead to enhanced levels of star formation, a process that occurs in both the pre-coalescence (pair) phase \citep{Ellison2008, Patton2013, Scott2014} as well as in post-mergers \citep{Ellison2013, Bickley2022, Vasquez-Bustos2025}.  These stars count neither as \textit{in-situ} (a term which is usually considered to measure secular star formation) nor \textit{ex-situ} (accreted from an external source) and are truly `new' stellar mass spawned by the merger.   In the work presented here we aim to quantify this `new' stellar mass, sometimes referred to as the `burst mass fraction', that results from interaction-induced enhancements in the star formation rate.

\medskip

In order to measure the amount of mass in new stars created as a result of a merger, we must know both where and when to look.  The spatial location of merger-triggered star formation is predicted by simulations to be primarily central \citep[e.g][]{Cox2006, Moreno2015, Moreno2021}, although additional mechanisms such as shocks, tides and turbulent compression may also play a role at larger radii \citep[e.g][]{Barnes2004, Renaud2014, Renaud2022, Petersson2023}.  Early observations supported the prediction of centrally concentrated star formation in mergers, either through the measurement of blue centres \citep[e.g.][]{Ellison2010, Patton2011, Lambas2012} or the measurement of enhanced star formation rates (SFRs) within small apertures \citep[e.g. ][]{Barton2000, Lambas2003, Ellison2013}.  Integral field unit (IFU) observations have now firmly established the central nature of triggered star formation \citep{Barrera-Ballesteros2015, Pan2019, Thorp2019, Steffen2021}. However, the same spatially resolved observations have also demonstrated that starbursts (both merger-induced and secular) can elevate star formation rates out to large radii \citep[e.g. ][]{Cortijo2017,Ellison2018, Wang2019, Mun2024}. In a direct comparison between merger-driven and secular starbursts \citet{Thorp2024} recently showed that the former are more centrally peaked than the latter, but both have SFR enhancements out to at least 1.5 effective radii.  Taken together, these observations motivate the measurement of merger-induced mass growth both in the central regions \textit{and} in the extended galactic disk. 

\medskip

The question of `when' to assess the build-up of stellar mass is potentially more tricky.  Triggered star formation is known to begin well before coalescence, as demonstrated both in simulations \citep[e.g.][]{Perez2006, Cox2008, Torrey2012, Patton2020} and the presence of enhanced SFRs in the pair phase in observations \citep{Nikolic2004, Ellison2008, Barton2007, Patton2013}.   However, this is just the start of the process and any measurement of stellar mass build-up in the pair phase will not capture the full extent of the growth.  Indeed, observationally, SFRs seem to peak in the post-merger regime \citep{Ellison2013, Bickley2022}, but how long does this enhancement persist after coalescence?  Knowing the answer to this question is critical, so that we can measure the central stellar mass excess once the epoch of triggered star formation is completed and a full accounting of new mass production can be made.

\medskip

Previously, our only guidance for the timescale over which star formation rates remained enhanced came from simulations, which predicted that SFRs returned to normal $\sim$ 0.5 -- 1 Gyr after coalescence \citep{Hani2020, Schechter2025}.  However, recently we have obtained a direct observational measurement of the evolution of SFR enhancement as a function of time.  Thanks to machine vision informed predictions of time post-merger ($T_{\mathrm{PM}}$), \citet{Ferreira2025} have found that SFRs peak around the time of coalescence, but decline to `normal' values after $\sim$ 1 Gyr, in good agreement with simulations \citep[e.g. ][]{Schechter2025}.  In order to quantify the mass of new stars formed as a result of mergers we are thus informed to look in the centres of post-merger galaxies that have coalesced around 1 Gyr ago, since these will have recently completed their merger-induced star formation.

\medskip

Having established where and when we should measure the burst mass fraction, the final challenge is to identify a sample in which the measurement can be made.  As mentioned above, recent advances with machine vision techniques are now making it possible to predict the time before/after coalescence from galaxy morphologies alone \citep{Koppula2021, Pearson2024, Pearson2025, Ferreira2025, Ferreira2026}.  Specifically, in the work presented here, we will use the $T_{\mathrm{PM}}$ predictions from the MUlti Model Merger Identifier \citep[\textsc{mummi},][Ellison et al. in prep;  Section \ref{mummi_sec}]{Ferreira2024, Ferreira2026} to identify post-mergers that have recently completed their triggered star formation (Section \ref{pm_sec}).  Quantification of excess stellar mass is made via a comparison with a control sample matched in $M_{\star}$ and redshift (Section \ref{control_sec}).  Burst mass fractions are computed within fixed angular apertures (Section \ref{df_sec}), within the bulge (Section \ref{db_sec}) and, for those galaxies observed with the MaNGA IFU survey, within fixed physical radii and annuli (Section \ref{dm_sec}).  We discuss our results in the context of previous observational measurements of the burst mass fraction in Section \ref{discuss_sec}.

\section{Data}\label{data_sec}

\subsection{A brief overview of \textsc{mummi}}\label{mummi_sec}

\textsc{mummi} combines an ensemble of 10 convolutional neural networks and 10 vision transformers to yield 20 independent votes on the status of a given galaxy image \citep{Ferreira2024}.  \textsc{mummi} is trained on labelled images of pre-coalescence pairs, post-mergers and isolated (non-mergers) generated from the Illustris TNG100-1 (hereafter, TNG) simulation \citep{Nelson2019}. The training set consists of galaxies above a mass limit of log M$_{\star} > 10^{10}$ \msun\ (to avoid limitations due to mass resolution in the simulations) and z$<$1. Mergers (both in the pair and post-coalescence regimes) have stellar mass ratios between 1:1 and 1:10.  \textsc{mummi} is therefore trained to identify both major and minor mergers, which are expected to trigger the most significant effects \citep[but see][for the non-negligible effects of `mini' mergers]{Bottrell24, Byrne25}. 

\medskip

Using simulations for the training set is appealing due to both the large number of available galaxy images and the ability to categorically know the ground truth about the merger status. It has been previously demonstrated that TNG well reproduces the large-scale morphologies of normal galaxies \citep{RodGom19, Zanisi21, Eisert24}, making it an appropriate choice for generating our mock images.  Although TNG (like all cosmological simulations) uses sub-grid recipes for small-scale processes such as star formation and AGN feedback, neural networks primarily identify mergers via their extended stellar tidal structures \citep{Cipri20, Gordon24}, whose morphologies (dominated by gravity) are expected to be well reproduced. Indeed, in their comparison between six different machine learning classifiers \citet{Marga24} found that runs trained on simulated images performed comparably to those trained on visually labelled mergers (at least in the mass and redshift range relevant for our work).  We also refer the reader to \citet{Ferreira2024} for details of the extensive testing specifically for the \textsc{mummi} pipeline. 

\medskip

\textsc{mummi} consists of three distinct classification steps.  In Step 1, \textsc{mummi} separates non-mergers and mergers (including both pairs and post-mergers), where the user defines the threshold number of votes (out of a maximum of 20) required for a positive merger classification.    Step 2 of \textsc{mummi} separates pairs from post-mergers.  Full details of the construction and training of Steps 1 and 2 of \textsc{mummi} can be found in \cite{Ferreira2024}.  Step 3 of \textsc{mummi} is applied only to post-mergers, and places each galaxy into one of four time post-merger ($T_{\mathrm{PM}}$) bins, up to a maximum of 1.76 Gyr beyond coalescence; see \cite{Ferreira2026} for more details. Taken together, the three stages of \textsc{mummi} are therefore able to not only separate non-mergers from pairs and post-mergers, but also assemble the latter category into a time sequence.

\medskip

It is well documented that observational realism is important for high performance during machine vision tasks \citep[e.g.][]{Bottrell, Wilkinson2024}.  The original \textsc{mummi} pipeline was trained using mock realism to match the $r$-band images of the Ultraviolet Near-infrared Optical North Survey \citep[UNIONS;][]{Gwyn2025}.  However, UNIONS $r$-band imaging is only available northwards of +30 degrees in declination.  In order to maximize the footprint to which \textsc{mummi} can be applied, it has subsequently been re-trained for application to the Dark Energy Camera Legacy Survey (DECaLS), providing coverage of more southern declinations (Ellison et al. in prep).  

\medskip

The parent catalog for both \textsc{mummi}-UNIONS and  \textsc{mummi}-DECaLS is based on the main galaxy sample of the Sloan Digital Sky Survey Data Release 7 \citep[SDSS DR7, ][]{Abazajian2009}, including galaxies in the redshift range $0.01 < z < 0.3$.  In order to maximize the post-merger sample used in the current work, we combine the two \textsc{mummi} post-merger catalogs, from UNIONS and DECaLS, as described in the next sub-section.

\subsection{Post-merger selection}\label{pm_sec}

Following our previous works \citep{Ferreira2025, Ellison2024, Ellison2025} we identify a galaxy as a post-merger if $>10/20$ votes from Step 1 of \textsc{mummi} are positive.  This majority vote fraction approach has been found to be a good trade-off between completeness and purity \citep{Ferreira2024}. We limit our observational sample to those galaxies with  log M$_{\star} > 10^{10}$ \msun, to reflect the sample on which \textsc{mummi} was trained.  Stellar masses are taken from the SDSS DR7 MPA/JHU catalog, as described in \cite{Kauffmann2003}\footnote{https://home.strw.leidenuniv.nl/$\sim$jarle/SDSS/}.

\medskip

In the work presented here, our objective is to assess whether there is long term central stellar mass growth as a result of galaxy mergers.  We therefore impose two further criteria on the post-merger sample.  First, we only use post-mergers in the longest $T_{\mathrm{PM}}$ bin, which corresponds to 0.96 $< T_{\mathrm{PM}} < 1.76$ Gyr.  The motivation here is to study galaxies after the merger-induced star formation has been completed. Since \cite{Ferreira2025} have shown that enhanced SFRs are statistically only found at $T_{\mathrm{PM}} < 0.96$ Gyr, galaxies in the final time bin should be no longer forming excess stars as a result of the merger.  Second, we select only post-mergers that are quenched.  This second criterion is adopted because we are, once again, interested in measuring the build-up of stellar mass after star formation is complete and there is no further growth.  We therefore impose the requirement that the specific SFR (sSFR) be less than log (sSFR/yr$^{-1}$) $< -11$ , where SFRs are again taken from the MPA/JHU catalog \citep{Brinchmann2004}. In practice, the longest post-merger time bin is dominated by quenched galaxies \citep{Ellison2024}, such that the majority ($\sim$70 \% at the lowest stellar masses and 95\% at the highest stellar masses) of galaxies are retained even with the application of this sSFR threshold.  As a result of these selection criteria we have a sample of 13,845 quenched post-mergers with log M$_{\star} > 10^{10}$ \msun\ and 0.96 $< T_{\mathrm{PM}} < 1.76$ Gyr\footnote{There are a handful of duplicates between the UNIONS and DECaLS surveys due to overlapping declination; these are removed from the final sample.}, of which approximately two thirds come from DECaLS.

\subsection{Control matching}\label{control_sec}

In order to identify a matched control sample for the post-mergers, we begin with a pool consisting of all non-merger galaxies in the UNIONS + DECaLS catalog, defined as having fewer than 3/20 positive votes in Step 1 of \textsc{mummi}. For each post-merger we search in turn for every galaxy in the non-merger pool that has the same stellar mass, SFR and redshift within tolerances of 0.05 dex, 0.05 dex and 0.005, respectively.  If fewer than five controls are found within these tolerances, we allow them to grow by a further 0.05 dex in both stellar mass and SFR and 0.005 in redshift.  The process is repeated a maximum of 5 times, for maximum tolerances in stellar mass, SFR and redshift of 0.25 dex, 0.25 dex and 0.025.  In practice, most post-mergers have many tens of matches without any adjustment to the original matching tolerances.  Only four post-mergers fail the matching completely and are removed from the sample.

\section{Results}

We present three complementary experiments to assess whether mergers result in central stellar mass growth.

\subsection{Enhancements in fibre stellar mass}\label{df_sec}

\begin{figure}
	\includegraphics[width=\columnwidth]{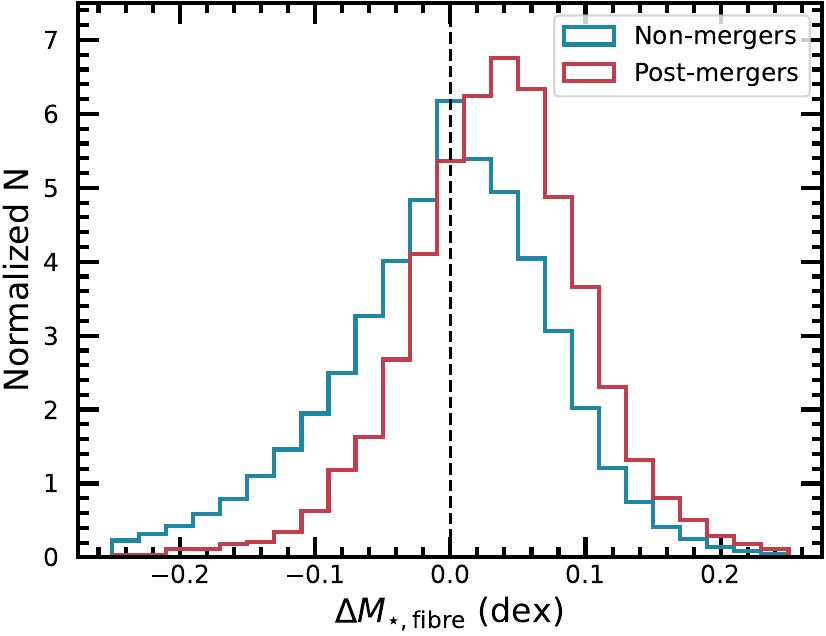}
        \caption{Normalized distribution of fibre stellar mass offset ($\Delta M_{\mathrm{\star,fibre}}$) for late-stage (0.96 $< T_{\mathrm{PM}} < 1.76$ Gyr) quenched post-mergers (red histogram) and non-mergers (blue).  The vertical dashed line is at zero. The median $\Delta M_{\mathrm{\star,fibre}}$ for the post-mergers is 0.04 dex, corresponding to a $\sim$ 10 \% enhancement. }
        \label{dmfibre}
\end{figure}

We begin our assessment of central stellar mass growth by making use of the fibre stellar masses that are provided in the MPA/JHU catalog.  The SDSS fibre has a 3 arcsecond diameter, corresponding to an approximate physical size of 1.5 kpc to 5 kpc for the redshift range 0.05 $< z <$ 0.2 over which most of our post-merger sample is located.  Despite this variable physical aperture, recall that our control sample is matched in both total stellar mass and redshift.  We therefore expect that the control sample for a given post-merger should have aperture coverage of the same physical and relative extent.

\medskip

We compute a fibre stellar mass `offset', $\Delta M_{\mathrm{\star,fibre}}$, as the difference (in log space) between the fibre stellar mass of a given post-merger and the median of its matched controls.  As a reference, and to quantify the underlying range in expected $\Delta M_{\mathrm{\star,fibre}}$ in the general galaxy population, it is also possible to compute the fibre mass offset for non-merger galaxies as well.  Figure \ref{dmfibre} shows the normalized distribution of $\Delta M_{\mathrm{\star,fibre}}$ for the late stage post-mergers (red) and non-mergers (blue).  As expected, the non-merger distribution is symmetric and centred around zero; the standard deviation of the distribution is 0.09 dex. However, for the sample of 13,841 late stage (0.96 $< T_{\mathrm{PM}} < 1.76$ Gyr) quenched post-mergers, the $\Delta M_{\mathrm{\star,fibre}}$ distribution is skewed towards positive values, with a median of 0.04 dex and standard deviation of 0.07 dex.  Although the magnitude of this median offset is not large (about 10\%), the distributions shown in Figure \ref{dmfibre} are clearly distinct and the signal of enhanced fibre mass is compelling\footnote{We note that although we have used the combined UNIONS + DECaLS catalogs from \textsc{mummi} for this work, we see an identical signal (median $\Delta M_{\mathrm{\star,fibre}}$= 0.04 dex) if we use just one of these catalogs, just with fewer statistics.}. A Kolmogorov-Smirnov (KS) test confirms that there is a zero percent probability that the two distributions are drawn from the same parent sample.

\begin{figure}
	\includegraphics[width=\columnwidth]{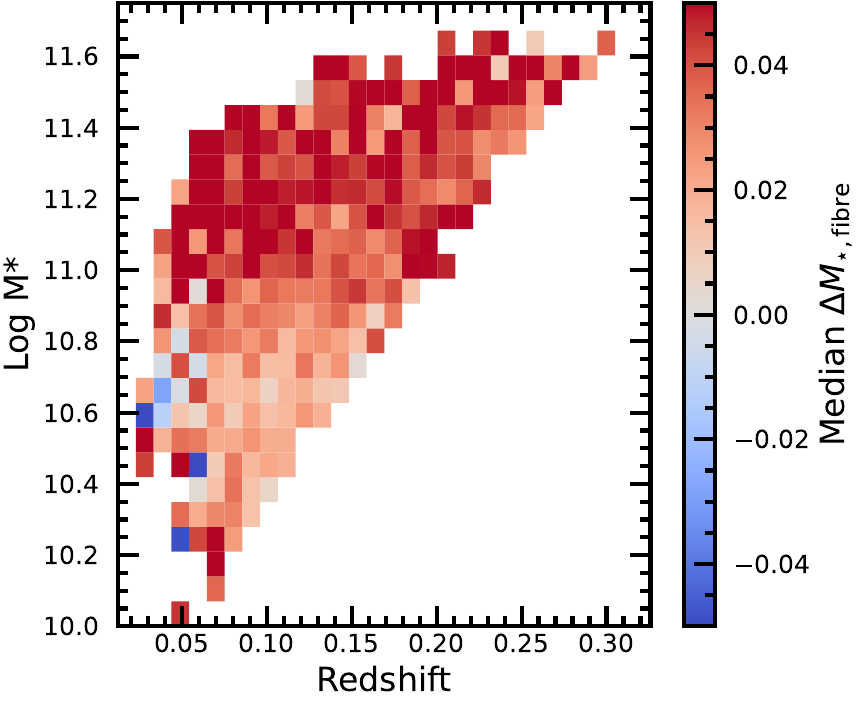}
        \caption{ The median values $\Delta M_{\mathrm{\star,fibre}}$ for the post-mergers computed in bins of redshift and stellar mass.  The enhancement in fibre stellar mass for post-mergers seen in Figure \ref{dmfibre} is present across the full stellar mass and redshift range of the sample. }
        \label{dmfibre_zm}
\end{figure}

Although the matching in stellar mass and redshift matching takes account of the variable coverage within the fixed 3 arcsecond fibre, Figure \ref{dmfibre} nonetheless contains masurements over a range of physical sizes. In order to assess the impact of varying redshift (and hence, varying physical coverage of the fibre), in Figure \ref{dmfibre_zm} we plot the median $\Delta M_{\mathrm{\star,fibre}}$ of the post-merger sample in bins of stellar mass and redshift.  Figure \ref{dmfibre_zm} shows that the enhancement in fibre stellar mass for post-mergers seen in Figure \ref{dmfibre} is present across the full stellar mass and redshift range of the sample.  However, we note that the enhancement in $\Delta M_{\mathrm{\star,fibre}}$ is slightly smaller (0.03 dex) at stellar masses log M$_{\star} < 10^{11}$ \msun.  The presence of a fibre mass enhancement even in the highest redshift post-mergers in the sample hints that the scale of the stellar mass growth extends to radii of at least $\sim$ 3 kpc.  In Section \ref{dm_sec} we will return to the question of the physical extent of stellar mass growth, using the overlap between MaNGA and \textsc{mummi}.

\subsection{Enhancements in bulge stellar mass}\label{db_sec}

In order to move beyond the fixed angular measurement offered by $\Delta M_{\mathrm{\star,fibre}}$, we now consider a more physically motivated metric.  Specifically, we make use of the bulge and disk masses presented in the catalog of \cite{Mendel2014}.   Bulge and disk masses are derived from spectral energy distribution fits to the SDSS optical magnitudes derived for these distinct morphological components by \citet{Simard11,Mendel2014}.  The two component photometry was obtained by modelling the five $ugriz$ bands as a de Vaucouleurs bulge plus exponential disk, where the scale lengths across each band were tied to the values in the $r$-band.  We refer the reader to \cite{Simard11, Mendel2014} for more details, including a full analysis of systematic and model uncertainties.

\medskip

Although galaxy mergers are typically characterized by their disturbed morphologies, our late stage post-merger sample shows only low surface brightness tidal features \citep[see examples in][]{Ferreira2026}.  Moreover, the central stellar bulge is still well-defined in all cases that we visually inspected.  Therefore, whilst measurements relating to the stellar disk may be somewhat affected by the interaction in our late-stage sample, the bulge fits (with a fixed Sersic index of $N$=4) should be robust.

\medskip

\begin{figure}
	\includegraphics[width=\columnwidth]{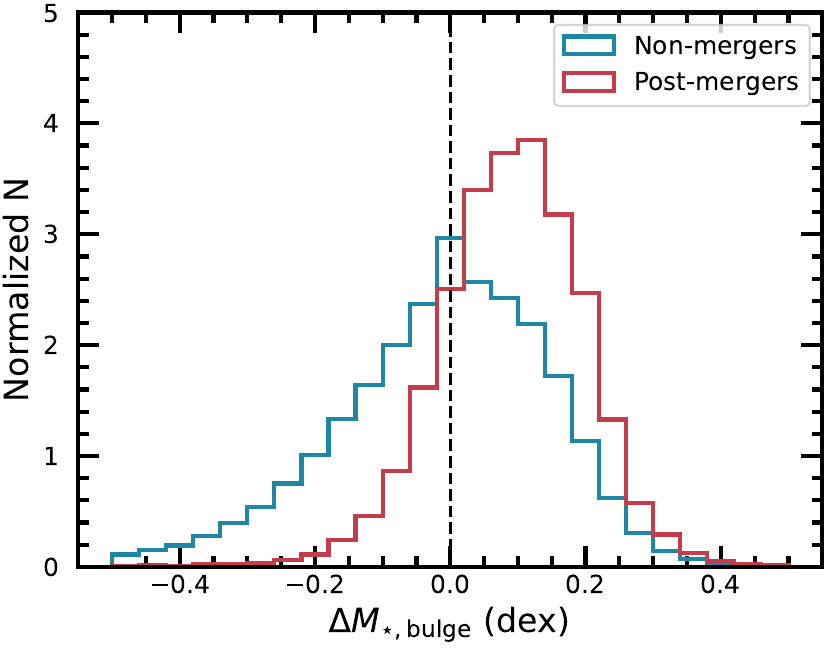}
        \caption{Normalized distribution of bulge stellar mass offset ($\Delta M_{\mathrm{\star,bulge}}$) for late-stage (0.96 $< T_{\mathrm{PM}} < 1.76$ Gyr) quenched post-mergers (red histogram) and non-mergers (blue).  The vertical dashed line is at zero. The median $\Delta M_{\mathrm{\star,bulge}}$ for the post-mergers is 0.09 dex, corresponding to a $\sim$ 20 \% enhancement.  }
        \label{dmbulge}
\end{figure}

A small fraction of our post-merger sample is not included in the Mendel catalog, reducing the sample from 13,845 to 13,506.  Control matching for this sample proceeds following the description in Section \ref{control_sec}, i.e. matching in total stellar mass, redshift and SFR.  The bulge mass offset ($\Delta M_{\mathrm{\star,bulge}}$) is then defined as the difference (in log space) between the bulge mass of a given post-merger and the median bulge mass of its matched control sample.  In Figure \ref{dmbulge} we show the distribution of $\Delta M_{\mathrm{\star,bulge}}$ for non-mergers (blue) and post-mergers (red).  As expected, the non-merger distribution is centred around zero; the standard deviation of the distribution is 0.15 dex.  However, the distribution of $\Delta M_{\mathrm{\star,bulge}}$ in the post-merger sample is skewed to positive values, with a median of 0.09 dex (with a standard deviation of 0.1 dex), corresponding to a $\sim$ 20\% enhancement.  A KS test confirms that there is a zero percent probability that the two distributions are drawn from the same parent sample.

\subsection{Radial enhancements in stellar mass using MaNGA}\label{dm_sec}

Our final test of central stellar mass growth in post-mergers utilizes the highest spatial resolution mass information at our disposal, namely the overlap between \textsc{mummi} and the MaNGA survey.  Of the $\sim$ 10,000 galaxies in the final MaNGA data release, there are 177 objects in common with our parent sample of 13,845 quenched post-mergers with log M$_{\star} > 10^{10}$ \msun\ and 0.96 $< T_{\mathrm{PM}} < 1.76$ Gyr.  However, in order to ensure adequate resolution for our spatially resolved tests, we further restrict the sample to the 61 targets at $z\le$0.05.  Given the median resolution of 2.5 arcseconds of the MaNGA survey, we therefore expect to achieve a physical resolution of at least $\sim$ 2 kpc.

\medskip

We use the publically released MaNGA DR7 data products for the work presented here.  Specifically, the stellar mass surface density (\sigstar), measured in each 0.5 arcsecond spaxel, is taken from the \textsc{pipe3d} catalog \citep{Sanchez2022}\footnote{Although the initial mass function (IMF) used in \textsc{pipe3d} differs from that used in both \cite{Mendel2014} and \cite{Kauffmann2003}, the differential nature of our mass analysis is agnostic to any choice of IMF.}.  In order to correct for inclination effects, we compute the de-projected galactocentric distance of each spaxel (in units of kpc), and also adopt inclination corrected \sigstar\ values throughout the analysis that follows.

\medskip

Not only is the size of our post-merger limited by the MaNGA overlap, but so too is the non-merger control pool, which contains $\sim$ 5800 galaxies.  We therefore adjust our control matching methodology to take just the five best matched non-merging control galaxies for each post-merger.  These best matched controls are assessed simultaneously in SFR, redshift, effective radius and stellar mass using a KDTree.  Due to the smaller dynamic range of redshift compared to SFR and stellar mass, the three variables are normalized before the matching process, so that they have equivalent weights.  As a result, the redshift, stellar mass and SFR are typically matched to within 0.001, 0.1 dex and 0.1 dex, respectively.

\medskip

\begin{figure*}
	\includegraphics[width=8.7cm]{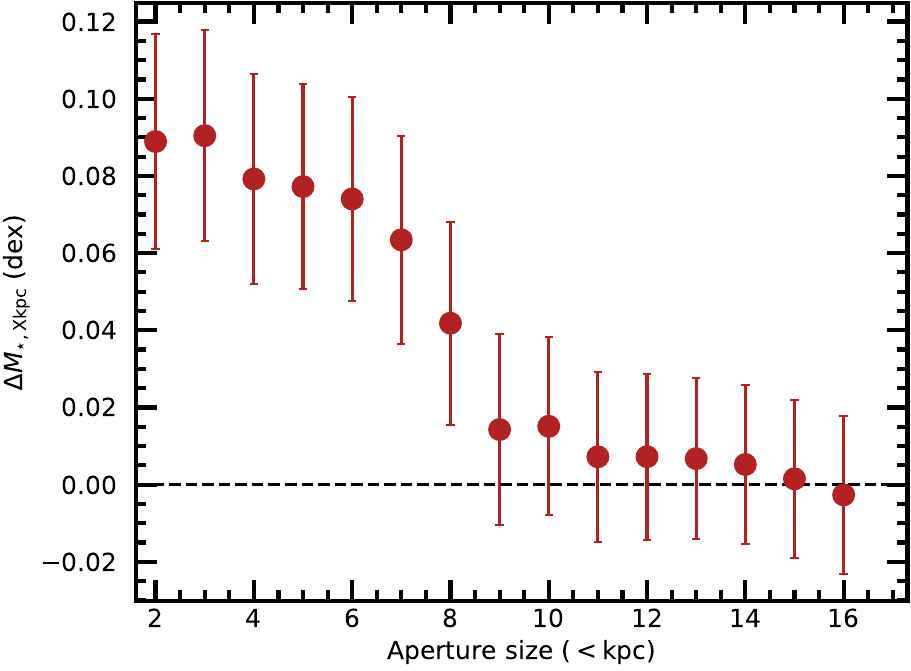}
	\includegraphics[width=8.7cm]{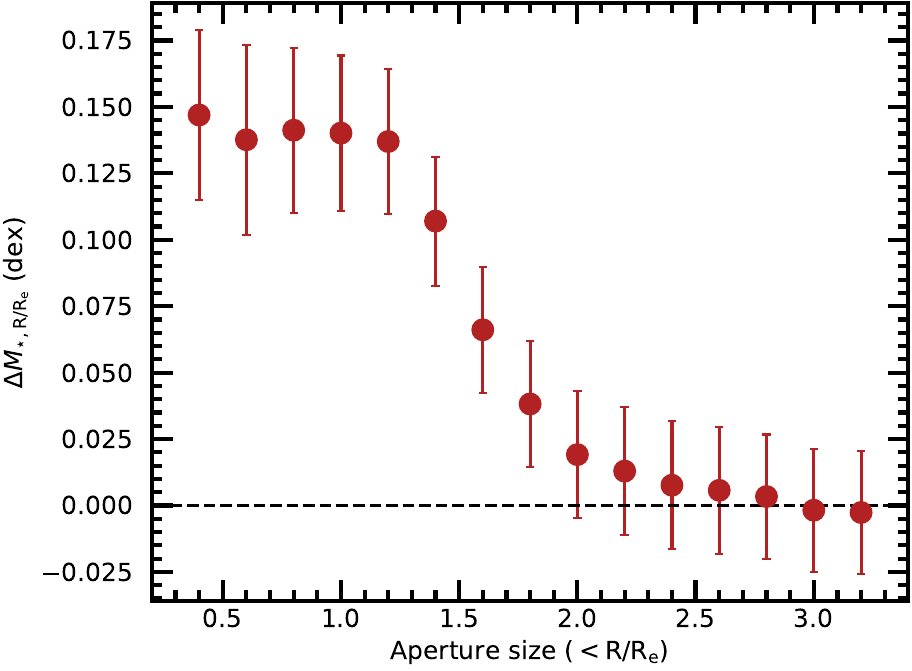}
        \caption{The excess mass measured using the 61 $z\le$0.05 quenched late-stage \textsc{mummi} post-mergers with MaNGA observations. The median mass excess is computed in apertures of increasing size relative to the five best matched control galaxies. Error bars are calculated as the standard error on the median.  The stellar mass excess is computed in radial apertures measured in units of kpc (left panel) and effective radius (right panel). When measured within a fixed kpc sized aperture we find an approximately 15 -- 20 \% mass enhancement within 7 kpc in the post-mergers.  When measured in apertures of fixed effective radius, we find a 40 \% mass enhancement with $\sim$ 1.25 R/R$_{\mathrm e}$.}
        \label{mass_rad}
\end{figure*}

For each post-merger and its five matched control galaxies, we sum \sigstar\ within fixed galactocentric distances (i.e. within apertures), repeating the calculation in radial bins defined in units of both kpc and effective radius.  To avoid any \textit{a priori} assumption of the scale on which central stellar mass growth occurs during the merger event, we compute masses in apertures with sizes from 2 up to 16 kpc (recall that, due to our redshift cut, we expect a physical resolution of at least $\sim$ 2 kpc) and up to 3 R/R$_\mathrm{e}$.   The stellar mass excess in the two sets of apertures, $\Delta M_{\mathrm{\star, X kpc}}$ and $\Delta M_{\mathrm{\star, R/R_e}}$, is then computed as the difference (in log space) between the summed stellar mass within $X$ kpc or R/R$_\mathrm{e}$ in the post-merger and the median value of the summed stellar masses within the same aperture of the matched control sample.  In Figure \ref{mass_rad} we plot the median stellar mass excess of post-mergers in bins of increasing aperture size, with uncertainties calculated as the standard error on the median.

\medskip

The left panel of Figure \ref{mass_rad} plots the mass excesses in kpc apertures. For apertures less than $\sim$ 7 kpc the typical $\Delta M_{\mathrm{\star, X kpc}} \sim$ 0.06 -- 0.08 dex, indicating an approximately 15 -- 20 \% stellar mass enhancement.   This offset decreases towards larger radii; for apertures larger than 11 kpc no offset is measured. The right panel of Figure \ref{mass_rad} plots the stellar mass excess in apertures defined by the effective radius.  A 40 \% stellar mass enhancement is seen within 1.25 effective radii, before decreasing for larger apertures until the control value is reached at $\sim$ 2.5 R/R$_\mathrm{e}$.

\medskip

\begin{figure}
	\includegraphics[width=\columnwidth]{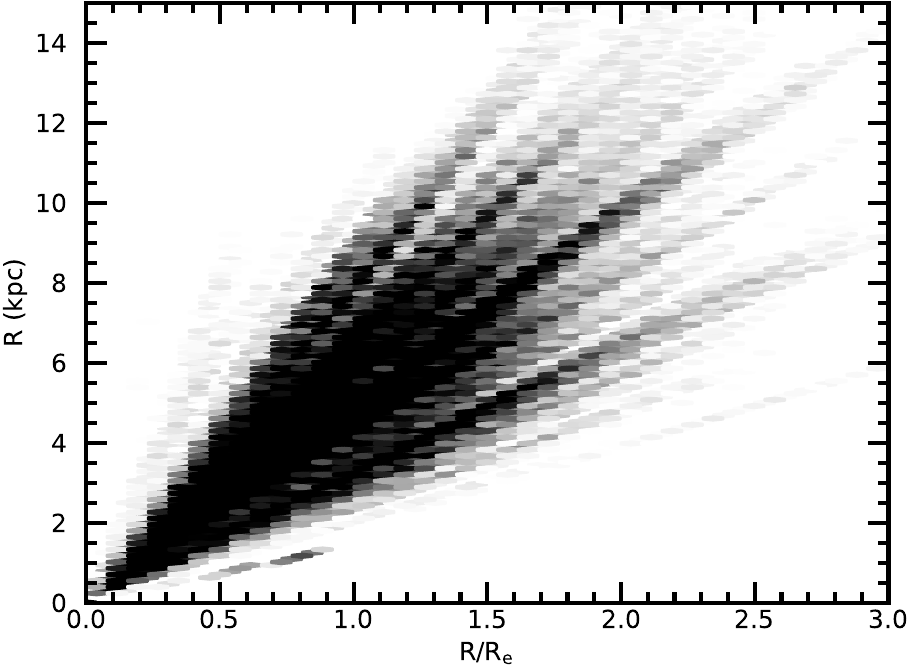}
        \caption{The relationship between galactocentric radius in units of kpc and effective radius for all spaxels in MaNGA in galaxies with \logm = 10.4$\pm$0.1.  There exists significant scatter, such that a fixed distance in kpc corresponds to an order of magnitude variation in effective radius.  }
        \label{kpc_aeff}
\end{figure}

In order to understand the different stellar mass enhancements in the left and right panels of Figure \ref{mass_rad}, in Figure \ref{kpc_aeff} we plot the relationship between the galactocentric radius of all spaxels in MaNGA as measured in kpc or units of effective radius, for galaxies in the limited total stellar mass range of \logm = 10.4$\pm$0.1.  It can be seen that a fixed kpc value corresponds to a wide range of R/R$_\mathrm{e}$.  For example, a spaxel at 5 kpc from the centre of its host galaxy could correspond to values of R/R$_\mathrm{e}$ as small as 0.3 or as large as 2.0.  One possible explanation of the different stellar mass enhancements as measured in apertures of fixed size in kpc vs. units of effective radius is that the latter is a better physical representation of the data, but the signal is blurred when using units of kpc.  This interpretation would suggest that the physical structure, at fixed stellar mass, plays a significant role in the dynamics of the gas flow and the distribution of the enhanced star formation.

\medskip

Regardless of the units used to define apertures, Figure \ref{mass_rad} hints that the stellar mass growth in late stage post-mergers is extended well beyond the central region.   Whilst measurements within apertures affords a maximum sample size (i.e. all 61 $z\le0.05$ MaNGA post-mergers have $\Delta M_{\mathrm{\star, X kpc}}$ and $\Delta M_{\mathrm{\star, R/R_e}}$ measurements for all apertures plotted in Figure \ref{mass_rad}), interpreting the exact scale of the merger induced stellar mass growth is non-trivial, because a positive signal within a given aperture can occur even if the actual growth is on a smaller scale.  We therefore repeat our calculation of $\Delta M_{\mathrm{\star, X kpc}}$ and $\Delta M_{\mathrm{\star, R/R_e}}$ but now use annuli instead of apertures.  Even though we have argued above that using radial bins in units of R$_\mathrm{e}$ is more sensitive to the underlying physical result, we have made our annulus calculation in both units for completeness. For the fixed kpc annuli we set the width of each annulus to be 2 kpc, which approximately matches the physical resolution of the MaNGA data for the low redshift cut we have imposed. The effective radius annuli are in increments of 0.5 R/R$_\mathrm{e}$, which is about 2.5 kpc for the typical galaxies in our sample (e.g. Figure \ref{kpc_aeff}).   Furthermore, we require that, for the calculation of $\Delta M_{\mathrm{\star, X kpc}}$ and $\Delta M_{\mathrm{\star, R/R_e}}$ in a given annulus, 1) the post-merger must have at least five spaxels with measured \sigstar\ and 2) at least three of its controls must also have measured \sigstar\ in that annulus.  If either of these criteria are not fulfilled, the post-merger is dropped from the sample. An increasing number of galaxies are lost as a result of insufficient spaxel coverage towards larger radii. For example, in the 9-11 and 11-13 kpc apertures only 38 and 25 of the original 61 post-mergers have MaNGA coverage, leading to larger measurement uncertainties at these radii.

\medskip

\begin{figure*}
	\includegraphics[width=8.7cm]{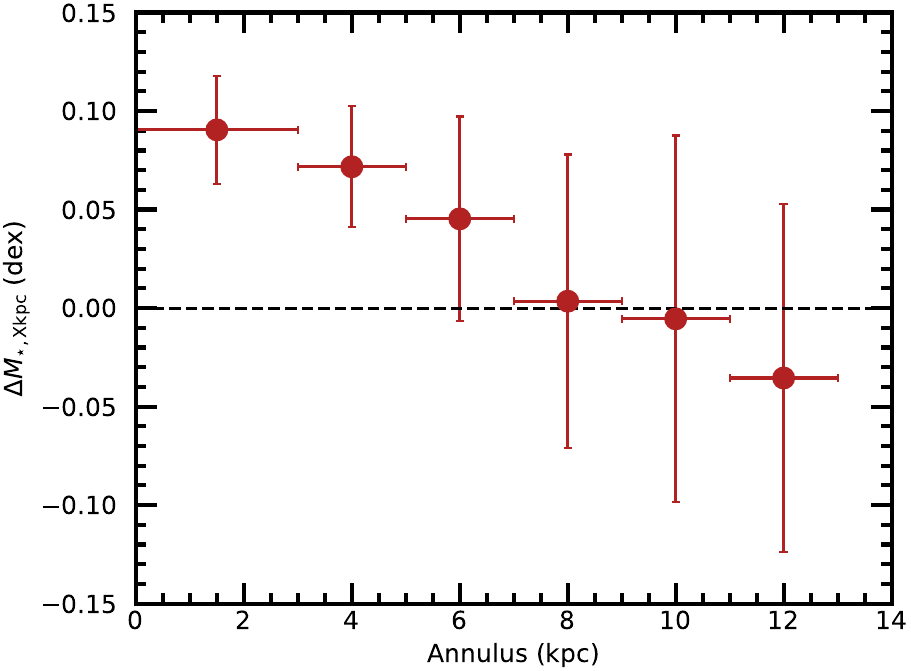}
	\includegraphics[width=8.7cm]{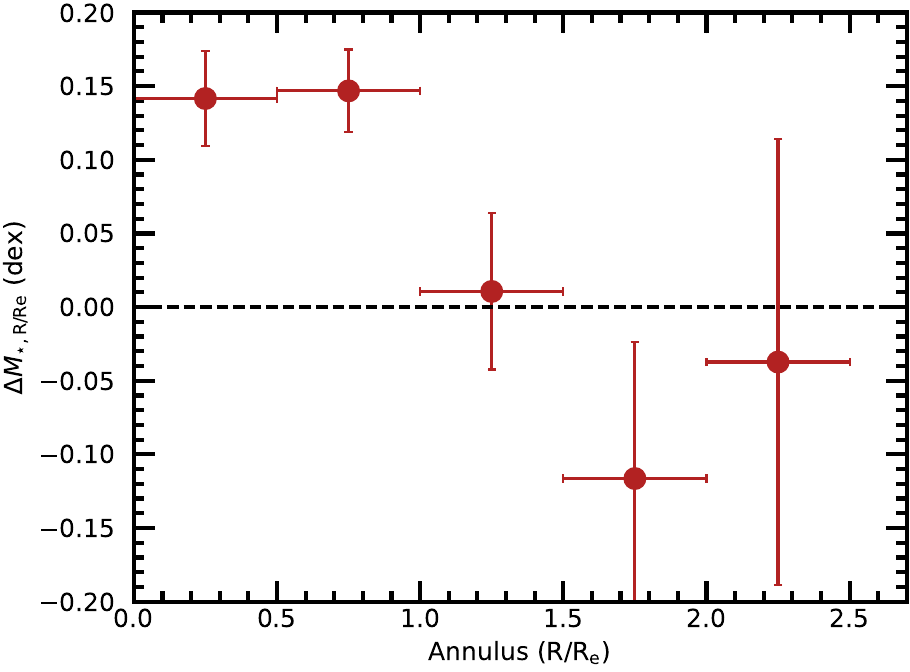}
        \caption{The median stellar mass excess measured in annuli in units of kpc (left panel) and effective radius (right panel). Error bars are calculated as the standard error on the median.   Significant stellar mass excesses are measured in annuli up to 7 kpc or one effective radius from the galaxy centre. }
        \label{mass_ann}
\end{figure*}

Figure \ref{mass_ann} shows the $\Delta M_{\mathrm{\star, X kpc}}$ (left panel) and $\Delta M_{\mathrm{\star, R/R_e}}$ (right panel) for our annulus calculations.  Positive stellar mass excesses with $\Delta M_{\mathrm{\star, X kpc}} \sim 0.07$ dex are measured from the centre out to a maximum annulus of 7 kpc. When plotting the annuli in bins of effective radius, the enhancement in $\Delta M_{\mathrm{\star, R/R_e}}$ is stable at 0.15 dex (40 \%) within one effective radius, with no statistical enhancement seen beyond this, although the small sample size limits the sampling of our radial bins.  Figure \ref{mass_ann} therefore confirms the interpretation of Figure \ref{mass_rad} that the decline in stellar mass enhancement in apertures exceeding $\sim$ 7 kpc or 1 R/R$_\mathrm{e}$ is caused by diluting the signal with regions experiencing no enhancement.

\medskip

The results presented in this sub-section, particularly those in annuli (right panel of Fig. \ref{mass_ann}) are obviously limited by the overlap between the \textsc{mummi} and MaNGA samples.  A much larger sample could be studied by using spatially resolved photometry.  For example, colour gradients would reveal complementary information to the study presented here on the location of recent star formation, and its evolution as a function of time through the merger sequence.  Such a photometric analysis is beyond the scope of the current paper, but will be presented by Woo et al. (in prep).

\section{Discussion and conclusions}\label{discuss_sec}

Galaxy mergers have long been known to experience elevated star formation rates \citep{Barton2007, Ellison2008, Patton2013}, and recent IFU surveys have confirmed that these enhancements are primarily centrally concentrated \citep{Barrera-Ballesteros2015, Pan2019, Thorp2019}, but with evidence of a spatially extended contribution out to at least 1.5 effective radii \citep[e.g. ][]{Cortijo2017,Ellison2018, Wang2019, Mun2024, Thorp2024}.  Assessing both the magnitude and location of the stellar mass that results from these starbursts is observationally challenging, and very few studies have reported quantitative measurements of the so-called `burst mass fraction'.  Here, we briefly review the main previous empirical attempts to measure this merger-induced mass growth, and place our own measurement in context.

\medskip

Early attempts to measure the burst mass fraction relied on quantifying cusps in the central light profile that departed from the $r^{1/4}$ law expected for the centres of merger remnants \citep{Mihos1994}.  Whilst analyses of merger light profiles established the presence of an `excess' of central light \citep[e.g.][]{Hibbard1999, Rothberg2004}, converting this to a mass fraction requires accurate modelling of the underlying disk profiles.  Through careful calibration with simulations, \cite{Hopkins2008} were able to determine a stellar mass growth of 3 -- 30 \%.

\medskip

The maturity of spectral fitting, derivation of star formation histories and forward modelling techniques has allowed a complementary approach to measuring merger-induced stellar mass growth.  Although the details of sample selection and model fitting vary, studies employing stellar population synthesis methods have found $\sim$ 10 -- 20 \% mass enhancement in merger remnants \citep{Cortijo2017,Yoon2023, Reeves2024}.

\medskip

Finally, large galaxy surveys are now approaching both the size and image quality where it might be possible to measure the additional mass created in mergers through an enhanced gravitational lensing signal.  \cite{Cheng2025} used $r$-band imaging from UNIONS \citep{Gwyn2025} coupled with the original \textsc{mummi} post-merger sample of \cite{Ferreira2024} to put an upper limit on the extra mass fraction at 60 \%.  Larger merger samples are evidently still needed in order to produce a measurable excess lensing signal, which will likely be forthcoming, e.g. with Euclid. 

\medskip

The ability of \textsc{mummi} to assign a timescale to a given post-merger has now opened up a further new route to measuring the \textit{in-situ} mass contribution from a merger-induced starburst, and one that is arguably much simpler than previous efforts.  By integrating under the curve of SFR enhancement vs. time since merger, \cite{Ferreira2025} determined that post-mergers with log M$_{\star}> 10^{10}$ \msun\ host 10--20 \% more stellar mass than their mass matched controls.  This method requires no modelling or choice of priors and has no intrinsic population degenaracies.  It does, however, only produce a population average result and also assumes that star formation rates follow smooth (rather than highly stochastic) trajectories. If, on the other hand, merger induced star formation is bursty and highly stochastic, interspersed with extended periods of `normal' SFRs, the 10 -- 20 \% burst mass fraction predicted by \cite{Ferreira2025} could be an over-estimate.

\medskip

In the work presented here, we have conducted a complementary analysis to the one presented in \cite{Ferreira2025}, using a sample of post-merger galaxies predicted by \textsc{mummi} to have recently completed their merger-induced burst.  However, we extend the sample of post-mergers used in \cite{Ferreira2025} by applying the \textsc{mummi} models to DECaLS $r$-band imaging, as well as the original UNIONS sample.  As a result, our post-merger sample contains almost 14,000 quenched, late-time post-mergers for which we conduct an measurements of stellar mass growth using SDSS imaging and spectroscopy.  Moreover, 61 of these post-mergers are in the final MaNGA data release at $z\le0.05$, allowing a more spatially resolved analysis.

\medskip

By measuring enhancements in the fibre stellar mass ($\Delta M_{\mathrm{\star,fibre}}$, Figure \ref{dmfibre}), bulge stellar mass ($\Delta M_{\mathrm{\star,bulge}}$, Figure \ref{dmbulge}), or within a fixed physical radius ($\Delta M_{\mathrm{\star,X kpc}}$ or $\Delta M_{\mathrm{\star,R/R_e}}$, Figure \ref{mass_rad}) we find burst mass fractions of 10 -- 40 \%.  This work extends that of \cite{Ferreira2025} by making a \textit{direct} measurement of excess stellar mass, and allowing us to do so on a galaxy-by-galaxy basis, without any assumption on its star formation history.  Whereas \citet{Ferreira2025} predict a 10--20 \% enhancement in stellar mass due to merger triggered star formation, our estimates (up to 40 \% in the central regions of post-mergers covered by MaNGA) are somewhat higher.  These differences are most likely rooted in methodological differences.  \citet{Ferreira2025} estimates the burst mass fraction based on the elevated SFRs measured in their merger sample.  However, the work presented here instead measures stellar mass directly, be it in the SDSS fibre, the bulge or in radial bins using MaNGA.  Our methodology is therefore sensitive to not only new stellar mass created as a result of the merger, but also to accreted stellar mass, i.e. the more traditional `\textit{ex-situ}' component.  The newly formed stars and those that have been accreted could potentially be disentangled through full spectral fitting of the MaNGA data cubes which could distinguish stellar ages.  

\medskip

The overlap between our \textsc{mummi} post-merger sample and the final data release of MaNGA enables us to further quantify not only the magnitude but also the physical extent of the stellar mass enhancement.  Merger-induced starbursts are frequently referred to as `central', and indeed simulations generally find enhancements within the central $\sim$ kpc \citep[e.g. ][]{Hopkins2008, Moreno2015, Moreno2021}.  It may therefore seem surprising that we measure an excess of stellar mass in post-merger remnants even within MaNGA annuli as large as 7 kpc or 1 R/R$_\mathrm{e}$ (Figure \ref{mass_ann}).  One possible explanation for this extended stellar mass excess is that some of the signal within these larger annuli could be classical \textit{ex-situ} mass, i.e. stars accreted from the merging companion that settle within the disk \citep[e.g.][]{Kannan15,Angeloudi2025}.  On the other hand, measurements of SFR enhancements in galaxy mergers using IFU samples find that, although the starburst is centrally concentrated, enhancements are seen throughout the disk \citep{Cortijo2017,Pan2019, Thorp2019, Thorp2024}. Our findings are completely consistent with these IFU-based measurements of SFR enhancement, without the need to invoke a contribution from accreted stars.  Moreover, using a sample of 31 post-mergers mapped in CO(2-1) with the KILOGAS survey (Davis et al., in prep), Ledger et al. (in prep) have found molecular gas enhancements also extend out to several kpc.  Taken together, the measurements of enhancements in molecular gas (Ledger et al., in prep), star formation rate \citep{Pan2019, Thorp2019, Thorp2024} and stellar mass (this work), paint a picture in which galaxy mergers lead to growth in both gas and stars over extents of at least the effective radius, significantly larger than is generally predicted in simulations.

\medskip

We have interpretted the excess stellar mass measured within the SDSS fibre, MaNGA apertures and annuli, and in the galactic bulge to be the result of star formation triggered during the merger that peaked $\sim$ 1 Gyr previously.  However, strictly speaking, the mass excesses we have measured in our sample of quenched post-mergers are relative to other quenched galaxies that have not had a recent merger, rather than their star-forming progenitors. From a dynamical perspective, the central mass concentration formed during a merger is expected to be long-lived.  Therefore, if most stellar mass in the bulges of (all) quenched galaxies (including those used in our control sample) are built from previous mergers, the enhancements we have measured might seem surprising.

\medskip  

In practice, massive, quenched galaxies are expected to assemble their mass from a combination of \textit{in-situ} star formation and mergers across a wide range of mass ratios.  Minor mergers are common \citep{FM08}, but tend to deposit their accreted stellar mass in the halo \citep{Karademir2019}, rather than in the regions we have detected mass enhancements.  Major mergers are relatively rare at low redshift, with only 10 - 50 \% of galaxies in the mass range of our sample expected to have had a major merger since $z\sim1$ \citep{Lin04,Lopez-SJ12}.  Even at high redshift, where the merger rate is elevated, major mergers contribute only a small fraction of the stellar mass \citep{Puskas25}.  Observations and simulations alike therefore indicate that the quenched galaxies that constitute our control sample are likely to have had their stellar mass built mainly through \textit{in-situ} growth \citep[e.g.,][]{Bundy09,Robaina2009, Martin17}.  Our results in turn imply that a recent major merger leaves an identifiable footprint on top of this history dominated by other growth mechanisms.

\medskip

In summary, the work presented here demonstrates that merger-triggered star formation can contribute an additional 40 \% to the stellar mass of a galaxy over galactocentric radii out to at least the galaxy's effective radius.

\acknowledgments

We are grateful to the two anonymous referees for insightful comments that improved the analysis and presentation of this work.

\medskip

\bibliographystyle{mnras}
\bibliography{bibliography}

\end{document}